\documentclass[doublecol]{epl2} 

\title{Nonlinear behavior of bosons in anisotropic optical lattices}

\author{A. Cetoli, E. Lundh}

\institute{Department of Physics, Ume{\aa} University, SE-90187 Ume\aa,
  Sweden}


\abstract{ We investigate the behavior of an array of
  Bose-Einstein condensate tubes described by means of a Bose-Hubbard
  Hamiltonian. Using an anisotropic non-polynomial Schr\"odinger
  equation we link the macroscopic parameters in the Bose-Hubbard
  Hamiltonian to the ones that are tunable in experiments. Using a
  mean field approach we predict that increasing the optical lattice
  strength along the direction of the tubes, the condensate can
  experience a reentrant transition between a Mott insulating phase
  and the superfluid one.}

\begin{document}
\maketitle

\section{Introduction}
Since Greiner {\it et al.} first succeeded at storing a Bose-Einstein
condensate in a two-dimensional array of narrow tubes using an optical
lattice \cite{greiner2001}, experimental progress on low-dimensional
Bose systems has been tremendous. To name but a few highlights, the
Mott transition was observed in three \cite{greiner2002}, one
\cite{stoferle2004}, and two dimensions \cite{kohl2005,spielman2007},
and in 1D a Tonks-Girardeau gas has been realized
\cite{paredes2004,kinoshita2005}. Moreover, in 2D the
Kosterlitz-Thouless transition was observed
\cite{hadzibabic2006,schweikhard2007}.  The dimensional crossover
between three, two, and one dimensions for bosons in an optical
lattice was studied theoretically using Tomonaga-Luttinger liquid (TLL)
theory in Refs.~\cite{ho,gangardt2006} and using Monte Carlo
simulations in Ref.~\cite{bergkvist2007}. These references studied 2D
or 3D optical lattices in which atoms can tunnel easily along one
Cartesian direction but not along the others. The system can then be
described as an array of tubes of bosons, which may or may not be
mutually phase coherent. As the tunneling between tubes is varied,
such a system will undergo a transition from a 3D superfluid (3D SF)
to a 2D Mott insulating phase (2D MI), which consists of a decoupled
array of 1D tubes. Similarly, if the tunneling probability along all
three Cartesian directions are made unequal, a 2D SF state can be
realized, in which there is superfluidity within separate 2D layers
but no coherence between them. These transitions are present only if
the tubes have finite length.

\revision{The theoretical approaches used in
  Refs.~\cite{ho,gangardt2006,bergkvist2007} are able to describe
  phase fluctuations within the tube-like filaments of bosons, but
  they do not capture any nonlinear effects due to the possible
  variation in the width of these tubes. Gross-Pitaevskii theory
  describes such variations \cite{oosten}; however, it does so at the
  expense of not being able to account for phase fluctuations.
  Keeping these limitations in mind, we offer in this Letter a
  description that is complementary to those of
  Refs.~\cite{ho,gangardt2006,bergkvist2007}, using Gross-Pitaevskii
  theory in order to understand how nonlinear effects may affect the
  phase transitions in this peculiar type of optical lattice.}
Such effects become important if the potential barriers
along the strongly coupled direction are so weak that each tube can be
considered as a quasi-1D Bose-Einstein condensate and the number of
bosons is large. It will be shown that such nonlinear effects can give
rise to a reentrant Mott transition in the array of 1D tubes.

The dilute Bose gas in an external potential $V_{ext}(\mathbf{r})$ is
described by the second quantized Hamiltonian
\begin{eqnarray}
\hat{H}[\hat\psi,\hat\psi^\dagger] = 
  \int d\mathbf{r}
   \Bigg[\frac{\hbar^2}{2 m}|\nabla \hat{\psi}(\mathbf{r})|^2
         &+& V_{\mathrm{ext}}(\mathbf{r})\,|\hat\psi(\mathbf{r})|^2 \\ \nonumber
	 &+& \frac{g}{2}\,|\hat\psi(\mathbf{r})|^4 
   \Bigg] \, .
\end{eqnarray}
The physical setting consists of a two-dimensional optical lattice in
the $x$ and $y$ plane with period $d_x$ and $d_y$, creating a square
array of tubes which develop along the $z$ direction.  Moreover a
weaker optical potential parallel to the tubes is added, so that in
the present case the external potential is given by
\begin{eqnarray} 
V_\mathrm{ext}(x,y,z) &=& - \frac{V_x}{2}\,\cos\frac{2 \pi \,x}{d_x}
                  - \frac{V_y}{2}\,\cos\frac{2 \pi \,y}{d_y}\\ \nonumber
               &+&\frac{V_z}{2}\,\cos\frac{2 \pi \,z}{d_z}
	       \, .  \label{extpot}
\end{eqnarray}

As in Ref.~\cite{jaksch} the many-body wavefunction of the gas
$\hat{\psi}(\mathbf{r})$ is rewritten as a sum of local operators
\begin{equation} \label{wannier}
\hat{\psi}(\mathbf{r})= \sum_{j_x j_y k} 
                        \phi_{\revision{j_x j_y} k}(\mathbf{r})\,\hat{b}_{jk} \, ,
\end{equation}

where $\phi_{j_x j_y k}(\mathbf{r})=\phi_k(x+j_x \,d_x,y+j_y\,d_y,z)$
and each $\hat{b}_{jk}$ acts on the $k^{th}$ state of the $j^{th}$
tube. The $\phi_{k}$ are a complete set of wavefunctions. In the
present setting we expect only the lowest energy state to be occupied,
so that it is possible to drop the $k$ index from~(\ref{wannier}).
Using expression~(\ref{wannier}) the second quantized grand potential
\begin{equation}
\hat{G}[\hat\psi,\hat\psi^\dagger]= 
  \hat{H}[\hat\psi,\hat\psi^\dagger] - \mu_{\mathrm{G}} \hat{N}[\hat\psi,\hat\psi^\dagger]
\end{equation}
can be rewritten
\begin{eqnarray}
\hat{G}
 = &-&\sum_{j} 
          \Big[ t_x   \, \hat{b}^\dagger_{j_x} \hat{b}_{j_x+1} 
                + t_y \, \hat{b}^\dagger_{j_y} \hat{b}_{j_y+1}
                + \mathrm{h.c.}\Big] \\ \nonumber
   &+& \frac{U}{2} \sum_j \hat{n}_j(\hat{n}_j-1)
   - \mu \sum_j \hat{n}_j 
   \, ,
\end{eqnarray}
where $j=(j_x,j_y)$, $\hat{n}_j=\hat{b}_j^\dagger \hat{b}_j$, and the
inter-tube tunneling matrix elements are
\begin{eqnarray}
t_\alpha= &-& \int d\mathbf{r} 
     \, \Big[ 
          (\mathbf{\nabla} \phi_{j_\alpha})^* 
            \cdot \mathbf{\nabla}\phi_{j_\alpha+1} \\ \nonumber
          &+& V_\mathrm{ext}(x,y,z)\, \phi_{j_\alpha}^*\,\phi_{j_\alpha + 1}
        \Big]
\, ,
\end{eqnarray}
with $\alpha=\{x,y\}$. The in-tube interaction energy is
\begin{equation}
  U= g\, \int d\mathbf{r} \, |\phi(\mathbf{r})|^4 \, ,
\end{equation}
\begin{equation}
\mu= - \int d\mathbf{r} \, 
            \left[\frac{\hbar^2}{2 m} |\mathbf{\nabla} \phi(\mathbf{r})|^2 
                           + |\phi(\mathbf{r})|^2 \, V_\mathrm{ext}(\mathbf{r})
            \right] + \mu_\mathrm{G} \, ,
\end{equation}
where $\mu_\mathrm{G}$ is a constant that fixes the number $N_\mathrm{tot}$ of
particles in the whole system and $g=4\pi\hbar^2 a_s/m$.

We assume that the external potential (\ref{extpot}) along the $x$ and
$y$ direction is very strong, so that the behavior of each tube can be
described by a one-dimensional approximation of the Gross-Pitaevskii
equation (cf. \cite{oosten}). This approach is not appropriate to
study the behavior of separate sites, since the number of particles
would be too small to get accurate results; it is however expected to
work in this context since each tube spans over many lattice sites.
The non-polynomial Schr\"odinger equation (NPSE) ~\cite{salasnich} has
proven itself to be an excellent means to describe such many
particle-systems. The main difference of this approach compared with
the TLL description \cite{ho} is that it is possible to consider
explicitly the transverse width of each tube, therefore studying the
nonlinear effects due to the finite density on the behavior of the
system. The main assumption of this work consists in taking the
groundstate of the NPSE as the Wannier function $\phi(x,y,z)$ that
appears in the expression of $t$, $U$, and $\mu$. In this way it is
possible to link the physical parameters that are tunable in a real
experiment to the macroscopic ones in the Bose-Hubbard Hamiltonian.
Notice that in general $V_x \ne V_y$, so an anisotropic version of the
NPSE is needed.

\section{Anisotropic NPSE}
We suppose that the confining potential in the transverse direction is
strong enough so that the behavior of each tube is well described
considering a harmonic external potential
\begin{eqnarray} 
  V_{\mathrm{tube}}(\mathbf{r})&\approx& \left[ \frac{m\,\omega_x^2}{2}\,x^2 
                                  + \frac{m\,\omega_y^2}{2}\,y^2\right]
                             + \frac{V_z}{2}\,
                                \cos\Big(\frac{2\pi}{d_z}\,z\Big) \nonumber\\
                     &=& \frac{m\,\omega_y^2}{2}
                             \left[ \Lambda \, x^2
                                    + y^2 \right]
                             + \frac{V_z}{2}\,\cos\Big(\frac{2\pi}{d_z}\,z\Big)
\, , \label{vtube}
\end{eqnarray}
where 
\begin{equation}
\omega_y= \frac{\sqrt{2}\pi}{d_y}\,\sqrt{\frac{V_y}{m}} \, ,
\end{equation}
and
\begin{equation}
\Lambda=\frac{\omega_x^2}{\omega_y^2}= \frac{V_x}{V_y} \, .
\end{equation}
Therefore the wavefunction of each tube is factorized as
\begin{equation} \label{ansatz}
\phi(\mathbf{r})= \frac{\Lambda^{\frac{1}{8}}}{\sqrt{\pi} \, \eta(z)}
                    \exp\Big[ -\frac{1}{2 \, \eta(z)}
                               \big(\sqrt{\Lambda}\,x^2 
                                 + y^2\big)\Big]\\
                    \times f(z) 
\, ,
\end{equation}
where 
\begin{equation}
\Lambda= \frac{V_x}{V_y} \, .
\end{equation}
\revision{We note that the approximation of the Wannier functions by
  Gaussians is known to introduce quite large quantitative errors when
  computing the tunneling matrix elements \cite{oosten}. Nevertheless,
  since the Gross-Pitaevskii equation does not allow an easy treatment
  using Bloch theory, it is very hard to improve on the Gaussian
  approximation.}

Following the steps in Ref.~\cite{salasnich} an equation for the
longitudinal part of each tube is derived. The energy functional for
the Bose-Einstein condensate in each tube is
\begin{eqnarray} \label{efun}
H[\phi,\phi^*]= \int d\mathbf{r} 
                 &\Big[& 
                     \frac{\hbar}{2 m}|\nabla\phi(\mathbf{r})|^2\\ \nonumber
                 &+& V_{\mathrm{tube}}(\mathbf{r})\,|\phi(\mathbf{r})|^2 \\ \nonumber
                 &+& \frac{g}{2} \,|\phi(\mathbf{r})|^4 \Big]
\, .
\end{eqnarray}
Substituting (\ref{ansatz}) into (\ref{efun}) it is possible to carry
out the integration along $x$ and $y$. Dropping the terms proportional
to $\eta'(z)$ \revision{\cite{salasnich2}} and minimizing with respect
to $\eta$ and $f^*$ the following equations are obtained:
\begin{eqnarray} \label{npse_code}
\Big[&-&\frac{\hbar}{2\,m}\frac{d^2}{dz^2}
   + \frac{V_z}{2}\, \cos \frac{2 \pi}{d_z} z \\ \nonumber
   &+& \,\revision{\frac{1+\sqrt{\Lambda}}{2}
     \,\hbar\,\omega_y}
     \,\frac{1 + \frac{3}{2}\gamma\,|f(z)|^2}
          {\sqrt{1 + \gamma\,|f(z)|^2}}
   \Big] \, f(z) \\ \nonumber
   &=& \mu\, f(z)
\, ,
\end{eqnarray}
\begin{eqnarray}\label{eta}
\eta_y(z)= a_y\, \left(1+\gamma\,|f(z)|^2\right)^{\frac{1}{4}} 
\, ,
\end{eqnarray}
where $a_y=\sqrt{\hbar/(m \omega_y)}$ and
\begin{equation}
  \gamma=4N\,a_s
                   \,\frac{\Lambda^\frac{1}{4}}{1+\sqrt{\Lambda}}
\, .
\end{equation}
Notice that Ref.~(\ref{npse_code}) reduces to the previously obtained
NPSE \cite{salasnich} in the case $\Lambda=1$.

\section{Bose-Hubbard Parameters}
Using~(\ref{ansatz}) as an \emph{ansatz} for the condensate
wavefunction it is possible to express $t_x$, $t_y$, $\mu$, and $U$ in
terms of integrals of $f(z)$ and $\eta(z)$. The expression for $t_y$
is
\begin{equation}
t_y = - \int dz \,\left[ \frac{\hbar^2}{2 m}\,A(z) 
                     + \frac{\hbar^2}{2 m}\,B(z) + C(z) \right]
\, ,
\end{equation}
where
\begin{eqnarray} \label{t1}
A(z)&=& \int dx\,dy 
            \Big[
              \Big(\frac{\partial \phi_{j_y}}{\partial x} \Big)^*
              \frac{\partial \phi_{j_y+1}}{\partial x} \nonumber\\
              &~& ~~~~~~~~~~~~~+
              \Big(\frac{\partial \phi_{j_y}}{\partial y} \Big)^*
              \frac{\partial \phi_{j_y+1}}{\partial y}
            \Big] \nonumber \\
      &=& \frac{\mathrm{exp}\left(-\frac{d_y^2}{4\, \eta_y(z)^2}\right)}
             {4 \eta_y(z)^2}\, |f(z)|^2 \nonumber \\ 
      &~& ~~\times\, \left( 2\,(1+\sqrt{\Lambda})\,\eta_y(z)^2 - d_y^2  
                             \right)
\, ,
\end{eqnarray}
\begin{eqnarray}
B(z) 
  &=& \int dx\,dy
      \left(\frac{\partial \phi_{j_y}}{\partial z} \right)^*
      \frac{\partial \phi_{j_y+1}}{\partial z}  \\ \nonumber
  &=& \mathrm{exp}\left(-\frac{d_y^2}{4\, \eta_y(z)^2}\right)
                \, |f'(z)|^2
\, ,
\end{eqnarray}
and
\begin{eqnarray}
C(z) 
  &=& \int dx\,dy \,V_\mathrm{ext}(x,y,z)\, \phi_{j_y}^*\,\phi_{j_y+1}\\ \nonumber
  &=&  \frac{1}{2}\,
                \mathrm{exp}\left(-\frac{d_y^2}{4 \, \eta_y(z)^2}\right) 
                \Bigg[
		  V_z\,\cos\left(\frac{2\pi\,z}{d_z}\right)\\ \nonumber
		&~&~~ - V_x \, \mathrm{exp}\left(-\frac{\pi^2 \eta_y(z)^2}
                                                   {\sqrt{\Lambda}\,d_x^2}
		                          \right) \\ \nonumber
		&~&~~ + V_y \, \mathrm{exp}\left(-\frac{\pi^2 \eta_y(z)^2}{d_y^2}
		                          \right)
          \Bigg]\, |f(z)|^2
\, .
\end{eqnarray}
\begin{figure}[tbp]
\includegraphics[width=3.4in, clip]{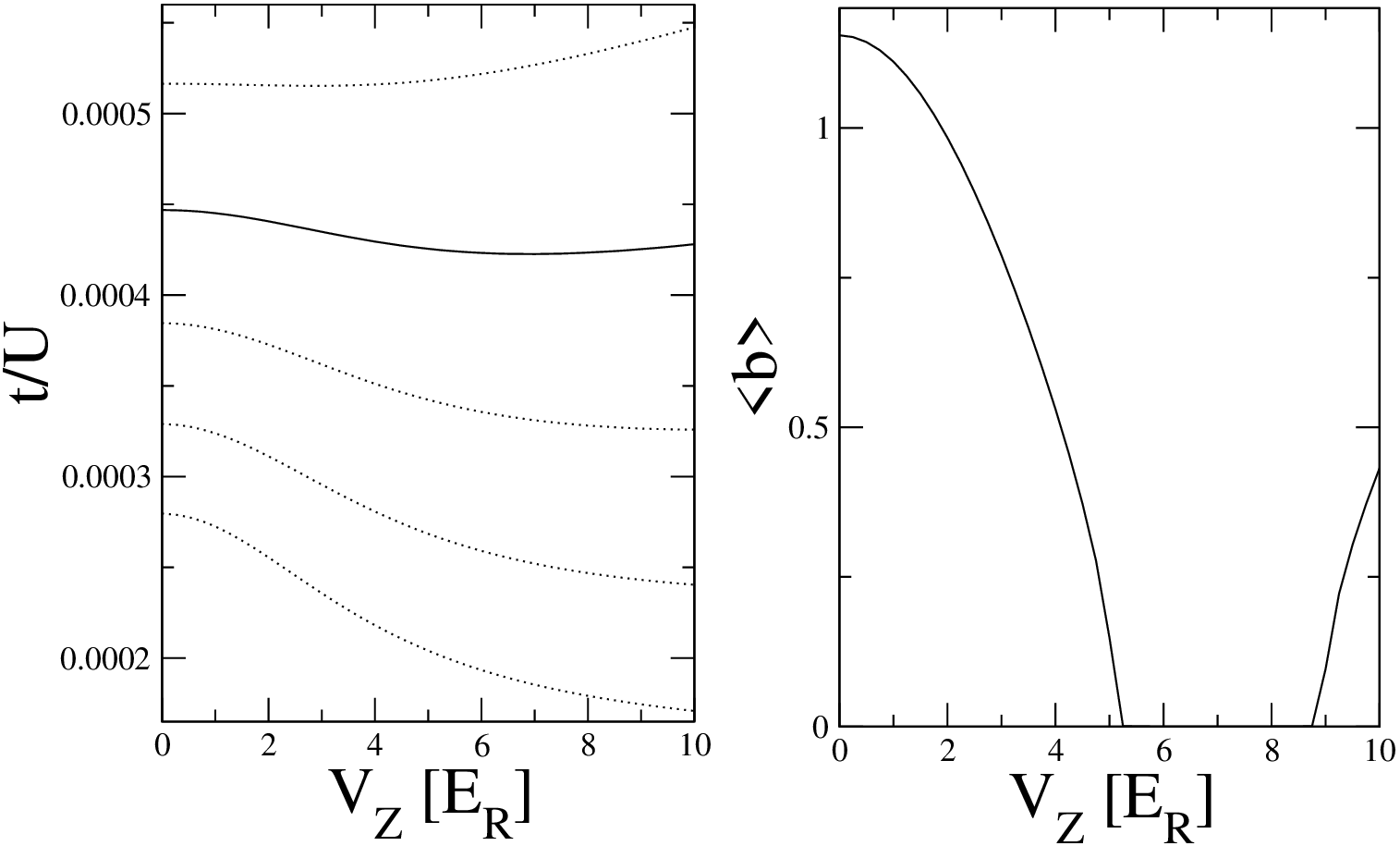}
\caption{Left panel: behavior of $t/U$ as a function of $V_z$ for
  $n_\mathrm{cell}=1,2,3,4,5$, while $V_x=V_y=26.4 E_\mathrm{R}$; for
  a system of length $L$ the parameter is scaled according to
  (\ref{t_on_U}).  Right panel: MF order parameter $\langle b \rangle$
  against $V_z$ for the case $n_\mathrm{cell}=4$ and $L=6$ ($2.55
  \mathrm{\mu m}$).}\label{bose_tubes_vz}
\end{figure}
Consistently with the NPSE approximation the terms proportional to
$\eta'(z)$ and $\eta'(z)^2$ have been omitted. The expression for
$t_x$ is obtained from $t_y$ switching the indices $x$ and $y$, and
changing $\Lambda$ to $\Lambda^{-1}$.  The integral for $U$ gives
\begin{eqnarray} \label{U}
U &=& g \int dx\,dy\,dz\, |\phi|^4 \\ \nonumber
  &=& g \,\bigg( \frac{V_x}{V_y} \bigg)^\frac{1}{4}
        \,\int dz \left[ \frac{|f(z)|^4}{2 \pi \, \eta_y(z)^2}\right]\, .
\end{eqnarray}
Switching the indices $x$ and $y$ leaves the value of $U$ invariant,
so that the expression for the self-interaction energy of each tube is
consistent with the symmetry of the problem.

In order to make explicit calculations we consider a system of
$^{87}$Rb atoms, whose atomic interaction is repulsive
($a_s=5.77\,\mathrm{nm}$). The lattice is a simple cubic one, with
$d_z=d_y=d_z= 425 \mathrm{nm}$.  Equation~(\ref{npse_code}) is solved
for $f(z)$ and $\eta(z)$ with periodic boundary conditions over a
single cell by means of a self-consistent approach. The length of the
tube affects the numerical results through the normalization of the
function $f(z)$, so that using periodic boundary conditions over $L$
cells the value of $t_\alpha$ ($\alpha=\{x,y\}$) is left invariant and
$U$ scales down as $1/L$, i.e.  $t_L=t$ while
\begin{equation}
U_L=\frac{U}{L} \,,
\end{equation}
where $t_L$ and $U_L$ denote the values of $t$ and $U$ for a system of
tubes with length $L$. Notice that
\begin{equation} \label{t_on_U}
\left( \frac{t_\alpha}{U} \right)_L = L \, \frac{t_\alpha}{U} \, .
\end{equation}
Denoting by $n_{\mathrm{cell}}$ the number of particles in each
longitudinal well, then the number of particles $n_{\mathrm{tube}}$ in
each tube is
\begin{equation} \label{n_ro}
n_{\mathrm{tube}} = L \, n_{\mathrm{cell}} \, .
\end{equation}
\begin{figure}[tbp]
\includegraphics[width=3.4in, clip]{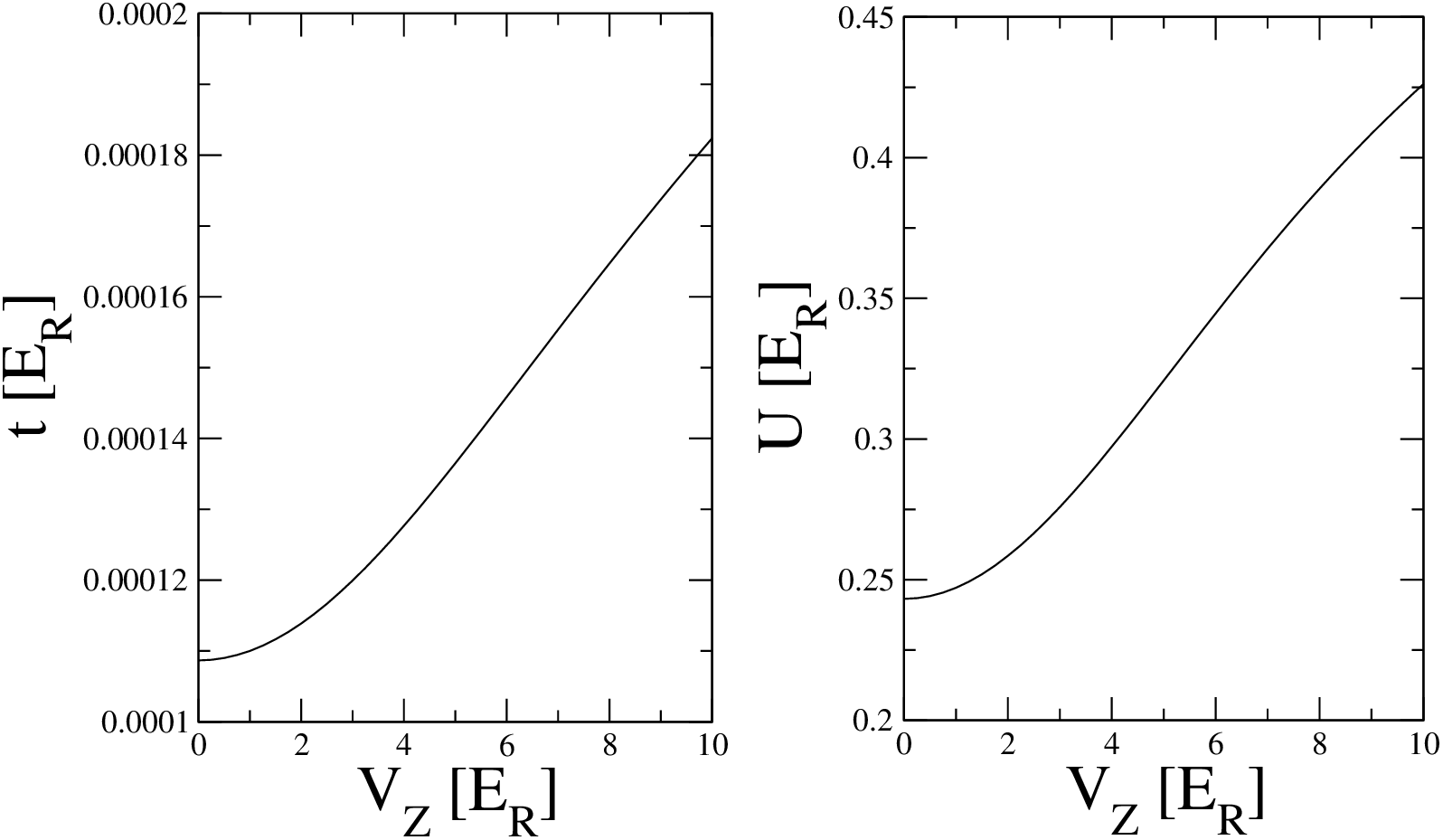}
\caption{Behavior of $t$ (left panel) and $U$ (right panel) as a
  function of $V_z$ for $n_\mathrm{cell}=4$, while $V_x=V_y=26.4
  E_\mathrm{R}$.}\label{bose_tubes_t_U}
\end{figure}
At first we focus on an isotropic case, i.e. a situation in which
$V_x=V_y=26.4 \,E_\mathrm{R}$, where $E_\mathrm{R}$ is the recoil
energy for a wavelength of 780 nm. Defining $t=t_x=t_y$, in
Fig.~\ref{bose_tubes_vz} (left panel) there appears the behavior of
$t/U$ while varying $V_z$ from $0$ to $10\,E_\mathrm{R}$, for
$n_\mathrm{cell}=1,2,3,4,5$.  According to Fig.~\ref{bose_tubes_vz},
for $n_\mathrm{cell}=4$ the ratio $t/U$ decreases until $V_z\approx 7
\,E_\mathrm{R}$, then it begins to rise. It is known
\cite{fisher,sachdev} that the Mott insulating phase appears as a set
of ``lobes'' in the plane $\mu/U$-$t/U$, each one corresponding to a
precise number of particles $n_{\mathrm{tube}}$ for each tube; outside
of the 2D MI zone the system moves along lines of constant
$n_{\mathrm{tube}}$. The initial decrease and subsequent decrease of
$t/U$ therefore means that for an appropriate length of the tubes the
system can cross the 3D SF - 2D MI transition twice. In order to give
an estimate of the critical potential strength a mean field (MF)
approach is employed (see for example \cite{georges, sachdev}). The MF
order parameter is given by the expectation value of the destruction
operator $\langle b \rangle$.  Figure~\ref{bose_tubes_vz} (right
panel) plots the behavior of this parameter for $n_{\mathrm{cell}}=4$
and $L=6$ ($2.55 \mathrm{\mu m}$).  The bounded interval in which the
superfluid phase vanishes is a clear consequence of the nonlinear
behavior of the system.

In order to understand the physical meaning of this phenomenon, the
plot in Fig.~\ref{bose_tubes_t_U} shows the behavior of $t$ and $U$
against $V_z$ separately for the case $n_\mathrm{cell}=4$. Increasing
the longitudinal optical lattice $f(z)$ becomes narrower, thus raising
the value of $U$ (\ref{U}), while at the same time the wavefunction
widens in the radial direction, increasing the tunneling rate $t$
(\ref{t1}).  At first $U$ rises faster than $t$ but for $t\approx 7
\,E_\mathrm{R}$ this relation is reversed.  The observed nonlinear
effect is therefore the result of a competition between $t$ and $U$.
\begin{figure}[tbp]
\includegraphics[width=3.4in, clip]{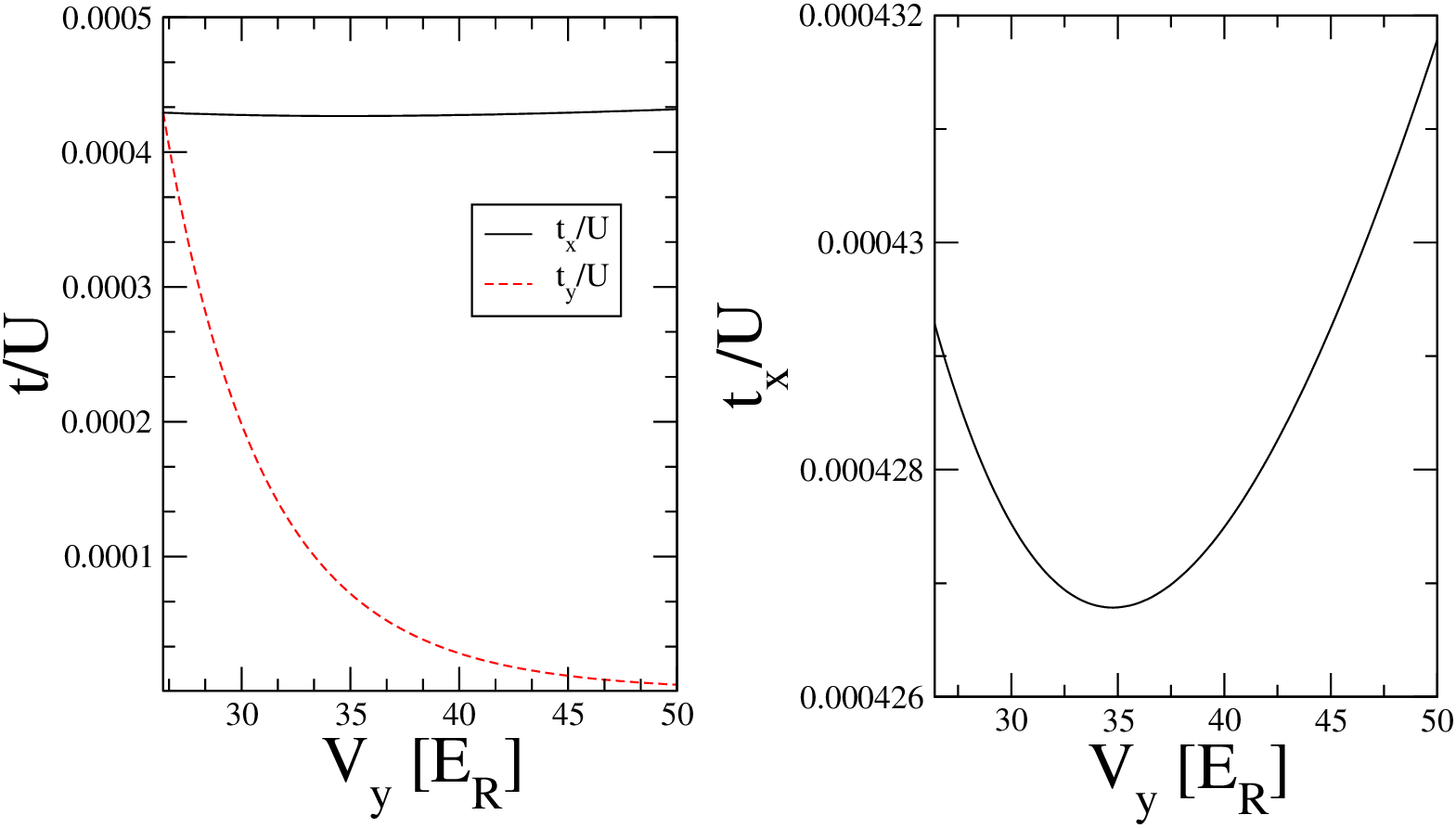}
\caption{Left panel: plot of $t_x/U$ and $t_y/U$ as a function of $V_y$,
  while $V_x= 26.4\,E_\mathrm{R}$, $V_z = 5\,E_\mathrm{R}$, and
  $n_\mathrm{cell}= 4$. Right panel: the behavior of $t_x/U$ is plotted
  in more detail for the same physical parameters.}\label{bose_tubes_tx_ty_vy}
\end{figure}
\begin{figure}[tbp]
\includegraphics[width=3.4in, clip]{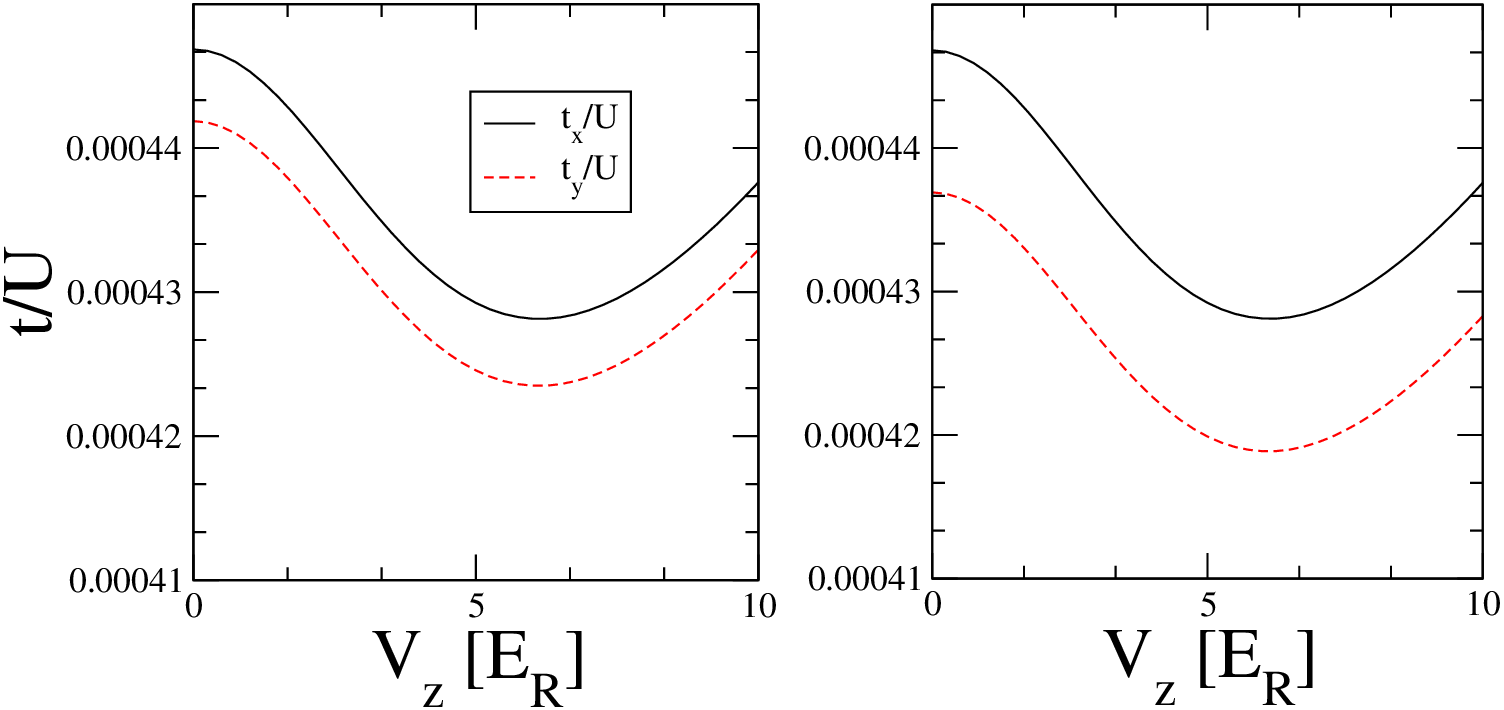}
\caption{Behavior of $t_x/U$ and $t_y/U$ as a function of $V_z$ while
  $V_x=26.4 \, E_\mathrm{R}$ and $n_\mathrm{cell}= 4$; left panel:
  $V_y=26.45 \, E_\mathrm{R}$, right panel: $V_y=26.5 \,
  E_\mathrm{R}$.}\label{bose_tubes_tx_ty_vz}
\end{figure}

So far it has been considered the isotropic case, in which $t_x$ is
identical to $t_y$. That is no longer true in an anisotropic setting,
i.e., one in which $V_x \ne V_y$. Figure~\ref{bose_tubes_tx_ty_vy}
(left panel) shows the behavior of $t_x$ and $t_y$ upon varying $V_y$,
with an occupation number $n_\mathrm{cell}=4$, keeping $V_x$ and $V_z$
fixed at $26.4\,E_\mathrm{R}$ and $5\,E_\mathrm{R}$ respectively. It
is shown that while $t_x$ varies very little for different values of
$V_y$, $t_y$ decreases exponentially. In the right panel of
Fig.~\ref{bose_tubes_tx_ty_vy} is plotted in detail $t_x$ against
$V_y$ for the same physical parameters, showing that $t_x$ has a
minimum for $V_y\approx 35\,E_\mathrm{R}$. These results suggest that
upon increasing $V_y$ the boson gas enters a 2D SF phase, in which the
system is organized in superfluid layers along the $y$ direction.
Moreover the behavior of $t_x$ suggests that - if all the parameters
are carefully tuned - these layers could experience a reentrant
transition between the 2D SF phase and the 2D MI one, in which each tube
is isolated from the others.

The asymmetric case provides another interesting phenomenon.
Figure~\ref{bose_tubes_tx_ty_vz} shows the situation in which
$V_x=26.4 \, E_\mathrm{R}$, and $n_\mathrm{cell}=4$, for $V_y=26.45 \,
E_\mathrm{R}$ (left panel) and $V_y=26.5 \, E_\mathrm{R}$ (right
panel). Increasing $V_y$, the curves of $t_x$ and $t_y$ become more
and more separated; in particular the curve for $t_y$ shifts
downwards, and its minimum moves to the left. This behavior suggests a
curious effect in which - upon increasing $V_z$ - the system may move
from a 3D SF phase to a 2D SF one, from this to a 2D MI phase and then
back again to the 2D SF.

\section{Conclusions}
In this Letter we have studied the behavior of a 2D array of strongly
elongated tubes of bosons in an optical lattice, employing the
Bose-Hubbard Hamiltonian in order to describe the 3D SF - 2D MI
transition.  Using the NPSE~(\ref{npse_code}) to compute the
groundstate in each tube it was possible to link the parameter $t$ and
$U$ of the Bose-Hubbard Hamiltonian to the physical parameters of the
atom gas.  Using this approach an observable nonlinear effect is
found: we predict that in an array of $^{87}$Rb tubes $2.55 \,
\mathrm{\mu m}$ long, with $d_x=d_y=d_z=425\,\mathrm{nm}$,
$V_x=V_y=26.4\,E_\mathrm{R}$, and an occupation number of 4 atoms for
each cell, the system goes through the 3D SF - 2D MI transition twice
while increasing $V_z$. The system itself is found to be in the
insulating phase for $V_z$ between $\sim 5 \, E_R$ and $\sim 8\,E_R$.
In addition, in the anisotropic case, our
results suggest that the nonlinear behavior of the system should cause
a reentrant transition between the 2D SF phase, in which the gas is
organized in superfluid layers, and the 2D MI phase, where each tube
acts independently.
\revision{
  A real experiment will be complicated by the fact that not
  all the tubes would be equally long and the occupation number
  fluctuates.  Moreover, due to the approximations we have made, 
  the most severe being the assumption of a Gaussian profile for the
  basis functions and the neglect of quantum fluctuations in the
  phase, the exact numbers are expected to differ from our
  predictions.  We argue that the qualitative analysis should hold,
  i.e., an experiment should in certain parameter regimes show a dip
  in the phase coherence while varying $V_z$, since this feature
  depends crucially only on the fact that the tunneling $t$ and
  on-site interaction energy $U$ exhibit different functional
  dependencies on the width of the tube-like condensates. }

\acknowledgements A.C. wishes to thank Robert Saers, Luca Salasnich
and Flavio Toigo for all the interesting discussions and precious
suggestions.

\end{document}